\documentclass[aps,pra, 
]{revtex4}
\usepackage{epsfig}
\usepackage{graphicx}

\newcommand{\eqb}{\begin{eqnarray}}
\newcommand{\eqe}{\end{eqnarray}}

\newcommand{\bv}{{\bf v}}

\newcommand{\br}{{\bf r}}

\newcommand{\jin}{j_\infty}

\newcommand{\rhin}{\rho_\infty}

\begin{document}

\title{Stationary vortex flows and macroscopic Zeno effect in Bose-Einstein condensates with localized dissipation}

\author{Dmitry A. Zezyulin}
\author{Vladimir V. Konotop}
\affiliation{
Centro de F\'isica Te\'orica e Computacional
Faculdade de Ci\^encias, Universidade de Lisboa, Avenida Professor Gama Pinto 2, Lisboa 1649-003, Portugal
and
Departamento de F\'isica, Faculdade de Ci\^encias, Universidade de Lisboa, Campo Grande, Ed. C8,
Lisboa 1749-016,  Portugal.
}

\begin{abstract}
We theoretically demonstrate a possibility to observe the  macroscopic Zeno effect in an effectively two-dimensional (pancake-shaped) repulsive Bose--Einstein condensate subjected to a strong narrow dissipation.  We show that the dissipation can generate stable stationary nonlinear flows which bear either zero or non-zero topological charge (vorticity). The superfluid flows towards the dissipative defect compensate the atomic losses. The macroscopic Zeno effect manifests itself in a nonmonotonous dependence of the  inward current density    on the strength of the dissipation.
\end{abstract}

\maketitle

\section{Introduction}

Bose-Einstein condensates (BEC), as one of a few quantum phenomena directly observable at a macroscopic scale, allow for detailed study of the relation between micro and macro worlds. In the mean-field approximation, a BEC is described by the Gross-Pitaevskii equation (GPE)~\cite{Pethick,Bose} for the macroscopic wavefunction, which accounts for the external trap potential and for the inter-atomic interactions, the latter giving origin to the nonlinearity in the GPE. This is a conservative model which  describes   well many experimental setups if the losses are negligible.

In the meantime, in many practical situations, losses are naturally present and (or) cannot be avoided. In particular, inelastic light-atom interactions~\cite{Pethick} and  inelastic two- and three-body inter-atomic interactions~\cite{two-body} lead to linear and nonlinear dissipative terms in the meanfield description. Emergent evolution of a BEC in linear and
nonlinear dissipative lattices was studies in~\cite{Abdul} and \cite{BludKon}, respectively.

One can consider condensates subjected to even stronger losses which, on the one hand cannot be treated as a perturbation, and, on the other hand, appear to be a useful tool for controlling and manipulating the system. Destruction of quantum states due to measurements, probing, or ionization by means of scattering are   examples of sources of relatively strong losses. One of the consequences of the respective ``dissipative'' evolution of   quantum states is their non-exponential decay, known since the pioneering works~\cite{Khalfin},  and even total suppression of the decay under continuous measurement~\cite{Zeno}. The latter phenomenon is known as the {\em quantum Zeno effect} (for a review see~\cite{review}). For the first time, the quantum Zeno effect was observed experimentally in vapors of Beryllium atoms~\cite{Itano}. In subsequent experimental studies, the phenomenon was observed by measuring escape of cold atoms loaded in an accelerating optical lattice~\cite{Zeno_accel_lattice}, exploring atomic spin motion controlled by circularly polarized light~\cite{Zeno_spin}, studying an externally driven mixture of two hyperfine states of a $^{87}$Rb BEC~\cite{Zeno_mixture}, and inducing production of cold molecular gases in an optical lattice~\cite{Zeno_molecule}. On the other hand, the quantum Zeno effect has also  received  considerable attention in theoretical works~\cite{theoretical}.

More recently, it has been suggested in~\cite{KS} that macroscopic (phenomenological) analog of the Zeno effect can  be addressed using a mean-field model of a BEC in a double-well trap with removal of atoms from one of the potential  wells. Comparison of the quantum simulations with the respective GPEs with a dissipative term has shown very good agreement, which immediately suggested that the phenomenon can be observed in a wider class of physical systems, such as, for instance, optical waveguides with Kerr nonlinearity~\cite{AKS}, whose dynamics is described by the equations fully analogous to the GPE with a double-well trap. Such a macroscopic manifestation of the phenomenon is referred to as the {\em macroscopic Zeno effect}.  

The macroscopic Zeno effect  is particularly well adjusted for some  experimental setting, such as probing of a BEC with an electron beam~\cite{GUHO,Ott}, where the condensed atoms are eliminated through the ionization resulting from electron-atom collisions. The macroscopic Zeno effect manifests itself in a nonmonotonic dependency of  the atomic losses on  the beam intensity (i.e.   on the effective dissipation in terms of the mean-field approximation). Very recently, the macroscopic Zeno effect was observed experimentally~\cite{Ott-experiment}.

The nonlinear dynamics in the presence of a localized defect is discussed already for many years; see e.g.~\cite{KM} where interaction of kinks of the driven long Josephson junction with a dissipative defect in a form of Dirac delta was considered. Turning to the effect of localized dissipative defects on the evolution of a BEC reveals numerous striking phenomena. Among them we mention a possibility of manipulating the condensate  {(say performing  switching and  phase locking)}~\cite{KS} or excitation of different macroscopic patterns in the condensates~\cite{BKPO}. It was also suggested that a strong enough dissipation can support quasi-stationary patterns~\cite{BKPO} and stationary currents~\cite{ZKBO} which depend on the intensity and the width of the electron beam   in a very nontrivial (nonmonotonic) way. In~\cite{SK2} it was shown that the macroscopic Zeno effect  can be observed not only in real space, but also in the Fourier one: a periodic dissipation (dissipative lattice) applied to a condensate allows for generating nondecaying Bloch states  {(see also \cite{FS})}.

Most of the previous studies  and, in particular, the theory of the macroscopic Zeno effect developed in~\cite{ZKBO}, however, dealt with quasi-one-dimensional settings.  Analysis of the phenomenon in the two-dimensional (2D) setting is the main   goal of this work.  More specifically, we  will show that a narrow  2D dissipation can generate stable stationary nonlinear flows which  bear either zero or non-zero topological charge (vorticity). Such modes exists due to compensation of the dissipative losses by the incoming flux (in this context we mention a recent for ~\cite{Porras} where similar mechanism resulted in stationary 2D vortex light beams in a cubic-quintic medium with homogeneous nonlinear dissipation were reported). We also show that the current density of  the stationary flows depends on the strength of the dissipation in the nonmonotonous way, which is  a manifestation of the macroscopic Zeno effect.

\section{The model}

We describe the 2D macroscopic Zeno effect  effect using   a model which governs the macroscopic (meanfield) dynamics of a pancake-shaped BEC with a positive scattering length $a_s>0$ (say, of $^{87}$Rb atoms), subjected to  a narrow electronic beam similar to the setting of \cite{GUHO,Ott-experiment}. In particular, we consider the model where the beam radius is of order of $r_0\approx 100$~nm while the radial dimension of the trap is $a_\bot\approx 3\,\mu$m ($\omega_\bot=2\pi\times 500$~Hz). In this situation, $r_0/a_\bot\approx 1/30\ll 1$ and one can explore the model of a narrow dissipative defect. To this end, we use the 2D GPE \cite{KS,WTHKGW}
\begin{eqnarray}
\label{GPE}
i \frac{\partial \Psi}{\partial t}=- \nabla^2\Psi+ g|\Psi|^2\Psi-i\Gamma(\br)\Psi,
\end{eqnarray}
in the dimensionless variables  $\br = (x,y)$ , where $\hbar=1$ and $m=1/2$.
We keep the nonlinear coefficient $g>0$ as a free parameter for the sake of convenience; alternatively, it can be scaled out by   rescaling the amplitude of the wavefunction. The  function $\Gamma(\br)\geq 0$ describes effect of the external dissipation.

Generally speaking, the macroscopic Zeno effect is understood as a suppression of atomic losses subject to increasing strength of dissipation. If there is no pump of the atoms into condensate, then the BEC   looses the atoms
with the rate 	
\begin{eqnarray}
\frac{dN}{dt}=-\int \Gamma({\bf r}) |\Psi(t, \br)|^2 d{\bf r},
\end{eqnarray}
 where $N=\int |\Psi(t, \br)|^2 d{\bf r}$ in the number of atoms in the BEC, and  the macroscopic Zeno effect manifests  itself in decrease of the   decay rate as  the intensity of the localized dissipation grows. In this statement, the effect was observed experimentally~\cite{Ott-experiment}.

In our present study,  we pose a    slightly different problem.  We consider an idealized situation when the dissipation, localized at $\br=0$, is applied to an infinite sample of the BEC (i.e. number of atoms $N$ is infinite). Then the dissipation induces  a particle current directed from   infinity towards the center, i.e. towards the dissipative defect. The     density of the inward current can be  adjusted  to   compensate   exactly   effect of the losses, which means that the condensate enters a stationary regime, i.e. the density distribution of the BEC and the density of the current  do not depend on time $t$. In what follows, such   structures are referred to as stationary flows. It is clear that the  inward current density of the stationary flow, generally speaking, depends on the strength of the dissipation.  Then the macroscopic Zeno effect, if exists, will manifest itself through a nonmonotonous dependence of the   current density   on  the strength of dissipation.  Such a  statement of the problem is justified by the fact that the rate of atomic losses is proportional to the current density towards the defect necessary to compensate losses and to ensure stationary flows.

\section{Stationary 2D flows}

In order to formalize the  statement of the problem, we rewrite the GP-equations in the hydrodynamic form
 \begin{eqnarray}
 \label{hydro1}
 \frac 12 \frac{\partial n}{\partial t}+\nabla(n\bv)+\Gamma(\br)n=0
 \\
 \label{hydro2}
 \frac{\partial{\bv}}{\partial t}+\nabla\left({\bf v}^2+gn-\frac{\nabla^2\sqrt{n}}{\sqrt{n}}\right)=0
 \end{eqnarray}
where $n(t,\br)=|\Psi(t,\br)|^2$ is the local density, $\bv(t,\br)=\nabla\Theta(t,\br)$  is the superfluid velocity, and $\Theta(t,\br)=\arg \Psi(t,\br)$ is the phase of the macroscopic wavefunction.

 In what follows the consideration will be reduced to the radially symmetric dissipative defect, i.e. we consider   $\Gamma(\br)\equiv \Gamma(r)$ (where $r=|\br|$). Moreover we consider dissipation having a finite support, i.e.
 \begin{eqnarray}
 \label{eq:fin_support}
 \Gamma(r)\equiv 0, \quad \mbox{for $r>\ell$}
 \end{eqnarray}
 and centered at $\br=0$ (this form of the dissipative term models, in particular, the electronic beam used in~\cite{GUHO,Ott,Ott-experiment}). Respectively, $\ell$ will be referred to as a width of the dissipation defect. Then the strength of the dissipation is proportional either to the maximal amplitude  $\Gamma_0$ of the dissipation (in our case $\Gamma_0=\Gamma(0)$) or to the width $\ell$.

Radially symmetric dissipation allows us to look for two-dimensional radially symmetric stationary vortex solutions of GP equation (\ref{GPE}) for which $n(t,\br)\equiv\rho^2(r)$ and $\Theta(t,\br)\equiv-\mu t + q \varphi +\theta(r)$, with $\varphi$ being the azimuthal coordinate, the integer $q$ characterizing the topological charge (vorticity) of the stationary flow, and $\mu$ being the chemical potential, i.e.
\begin{equation}
\label{statflow}
\Psi(t, {\bf r})  = \psi(r)e^{-i\mu t + i q \varphi},\quad \psi(r) = \rho(r)  e^{i\theta(r)},
\end{equation}
We also introduce the radial component of the superfluid velocity $v(r)=d\theta (r)/dr $ and  note  that the chemical potential is given as $\mu = g\rhin^2$, where $\rhin=\lim_{r\to\infty}\rho(r)$. Then substituting the expression (\ref{statflow}) into the GPE (\ref{GPE}),  we obtain the following system of two equations:
\begin{eqnarray}
    \rho_{rr} + \frac{1}{r}\rho_r - \frac{j^2}{r^2\rho^3} + \left(\mu- \frac{m^2}{r^2}\right)\rho - g\rho^3 =0,
    \label{eq:ODEsysa}\\
    j_r + r\Gamma(r)\rho^2 = 0,
    \label{eq:ODEsysb}
\end{eqnarray}
where
\begin{eqnarray}
\label{j}
j(r) = rv(r)n(r)
\end{eqnarray}
 is the radial current density which is necessary to compensate  the losses induced by the dissipative term $\Gamma(\br)$.

We are interested in solutions (\ref{statflow}) that  correspond to   stationary flows directed from infinity towards the center $r=0$. We assume that the    atomic density $n(r)$  is uniform at infinity, i.e. $n(r)$ approaches a fixed constant $n_\infty$ as $r\to \infty$. Without loss of generality, in what follows we   consider $n_\infty=1$ (other values of   $n_\infty$ can be addressed by means of rescaling the nonlinear coefficient $g$).  Since $\Gamma(r)\equiv 0$ for $r>\ell$, the density of  the radial inward current $j(r)$ is    constant for large $r$:  $j(r)=j_\infty$ for $r\geq \ell$.  Then for any stationary flow the chemical potential is fixed as $\mu=gn_\infty$.

For any stationary state,   the current density $j_\infty$  is such that the inward flow compensates exactly the effect of the losses $\Gamma(r)$. Hence  we conclude that the macroscopic Zeno effect (if any)  manifests itself in  a  nonmonotonous  dependence of $j_\infty$ on  the amplitude $\Gamma_0$ or on the width $\ell$ of the dissipation.

Since   $\Gamma(r)$ is zero outside of the  bounded domain $r<\ell$, one finds that
at $r\to\infty$ the limiting value of the density is approached algebraically (notice that in the 1D case the respective asymptotic is exponential~\cite{ZKBO}):
\begin{eqnarray}
\label{eq:1/r2}
\rho(r) = \rho_\infty -
\frac{1}{2gr^2}\left(\frac{m^2}{\rho_\infty} + \frac{j_\infty^2}{\rho_\infty^5}\right)  - O\left(\frac{1}{r^{4}}\right),
\end{eqnarray}
where $\rho_\infty = \sqrt{n_\infty}$.
Behavior of the radial velocity as $r \to \infty$ can be described as
\begin{equation}
v(r) = \frac{j_\infty}{r\rho^2(r)} =
\frac{j_\infty}{\rho^2_\infty r} + \frac{j_\infty }{\rho^3_\infty g r^3}
 \left(\frac{m^2}{\rho_\infty}+ \frac{j_\infty^2}{\rho_\infty^5}\right)+ O\left(\frac{1}{r^{5}}\right).
\end{equation}
Respectively, as $r\to\infty$, the   behavior of  argument $\theta(r)$ of the stationary wavefunction $\psi(r)$ is  logarithmic in the leading order:
\begin{eqnarray}
\theta(r) \approx \theta(R) + \frac{\jin}{\rho_\infty^2}\ln\frac{r}{R},
\end{eqnarray}
where  $r\geq R \gg \ell$.

Turning to behavior of the solutions in the vicinity of the origin, i.e. in the limit $r\to 0$,   we show that if the topological charge is zero, i.e. if $q=0$, then  $\rho(0)\neq 0$ and $v(r)\sim r$.
Indeed, let us make the most general assumption that   in the leading order $\rho=\rho_0 r^\alpha$ and $v\sim r^{2\beta-1}$ where $\alpha$ and $\beta$ are, so far, arbitrary constants. Then, from (\ref{eq:ODEsysb}) in the leading order we obtain $r^{2(\alpha+\beta-1)}\sim\Gamma_0r^{2\alpha}$, i.e. $\beta=1$ and $v\sim r$. Substituting the obtained asymptotic for $v$ in  (\ref{eq:ODEsysa}), and requiring $\alpha$ to be different from zero, we obtain in the leading order that the three terms in (\ref{eq:ODEsysa}) are respectively of order of $r^{-2}$, $r^4$ and $(\rho_0^2r^{2\alpha}-\rho_\infty^2)$. These terms cannot cancel each other in the limit $r\to 0$, i.e. the assumption about nonzero $\alpha$ is contradictory, and in the leading order $\rho\approx \rho_0\neq 0$ at $r\to0$.
For nonzero topological charge, $q\ne0$, the asymptotic at $r\to 0$ reads $\rho(r)=r^q(\rho_0 + o(1))$,  which is typical for vortex-like solutions.

We conclude this section by emphasizing that chosen dissipation is smooth but  has a {\em finite} support $\ell>0$.  To understand that the width of the defect $\ell$ is a relevant parameter (for the 1D case this is was shown in~\cite{ZKBO}) for the observation of the Zeno effect, let us consider a stationary
solutions $n_t=0$. Integrating (\ref{hydro1}) over a disc domain, of the radius $\ell$, covering the defect region,  we obtain
\begin{eqnarray}
\int_{r< \ell} \nabla(n\bv) d^2{\br}=-\int_{r< \ell} \Gamma(r)n d^2{\br}
\end{eqnarray}
 Now from the Green's divergence theorem we compute (recall the definition (\ref{j}))
\begin{eqnarray}
2\pi\ell n(\ell)v(\ell)=2\pi j (\ell) =-\int_{r< \ell} \Gamma(r)n d^2{\br}
\end{eqnarray}
(recall that $v$ is the radial component of the velocity and we consider radially symmetric solutions $n(r)$ and $v(r)$).
Since we assumed that the dissipation has a finite support, Eq.~(\ref{eq:fin_support}), we have that $j(\ell)=j_\infty$.
Thus, if $n(r)$ is not "too" singular function at $r=0$, i.e. is bounded in the limit $r\to 0$, in the leading order we have that in the limit $\ell\to 0$
\begin{eqnarray}
\label{limit}
j_\infty\approx -\frac{n(\ell)}{2\pi}  \int_{r< \ell} \Gamma(r) d^2{\br}
\end{eqnarray}
In other words $|j_\infty|$ is proportional to the integral strength of the dissipation  $ \int_{r< \ell} \Gamma(r) d^2{\br}$ and it monotonously grows with $\Gamma_0$ (or $\ell$) at fixed $\ell$ (or $\Gamma_0$) meaning that in the limit $\ell \to0$ one cannot observe the macroscopic Zeno effect. Formula (\ref{limit}) reveals also that $|j_\infty|\propto \Gamma_0$ at $\Gamma_0\to 0$ and  $\ell$ fixed, and $|j_\infty|\propto \ell^{2+2q}$ at $\ell\to 0$ and $\Gamma_0$  fixed. These types of behavior we will observe in Fig.~\ref{fig-j(g)m=0} below (the upper and lower lines).

\section{Numerical study of the stationary currents}

For numerical study of stationary flows we choose the dissipation in the form
 \begin{eqnarray}
 \label{gamma}
 \Gamma(r)\equiv \left\{
 \begin{array}{ll}
 \Gamma_0\left(r^2/\ell^2 - 1\right)^2  &\mbox{for $r\leq \ell$},
 \\
 0 & \mbox{for $r>\ell$}.
 \end{array}
 \right.
 \end{eqnarray}
Then $\Gamma(r)$ is a smooth  function with  finite support.

Since experimental verification of the predicted effects would be only possible using stable stationary currents, for all the results reported below we tested the linear stability of the respective solution. To this end, we used the standard  expression for the perturbed solution
\begin{equation}
\Psi = e^{-i\mu t+iq\varphi}\left[\psi(r)+ a_\kappa(r)e^{i\omega t+iq\varphi} + b_\kappa^*(r)e^{-i\omega^* t-iq\varphi} \right],
\end{equation}
where  $a_\kappa$ and $b_\kappa$  characterize the small perturbation. After substituting this expression into (\ref{GPE}) and subsequent linearization around $a_\kappa$ and $b_\kappa$, one arrives at    a set of decoupled eigenvalue problems
\begin{equation}
    \label{eq:LinStabMatrix}
    \left(\begin{array}{cc}%
    L_{+,\kappa}& -g\psi^2\\%
     g(\psi^2)^* & -{L_{-,\kappa}}%
    \end{array}\right)
    \left(\begin{array}{c}
    a_\kappa\\ b_\kappa
    \end{array}\right)=\omega \left(\begin{array}{c}
    a_\kappa\\ b_\kappa
    \end{array}\right), \quad \kappa=0,1,\ldots,
\end{equation}
with
\begin{equation}
L_{\pm,\kappa} = \frac{d^2\ }{d r^2} +
\frac{1}{r}\frac{d\ }{d r} + g\rho_\infty^2 \pm
i\Gamma(r) - 2g|\psi|^2 - \frac{(\kappa \pm q)^2}{r^2}.
\end{equation}
If for some $\kappa=0,1,\ldots$ the spectrum of  problem (\ref{eq:LinStabMatrix}) contains   an eigenvalue $\omega$ with negative imaginary part, then the corresponding stationary flow is unstable. Otherwise, the flow is linearly stable.

\begin{figure}
\includegraphics[width=0.75\textwidth]{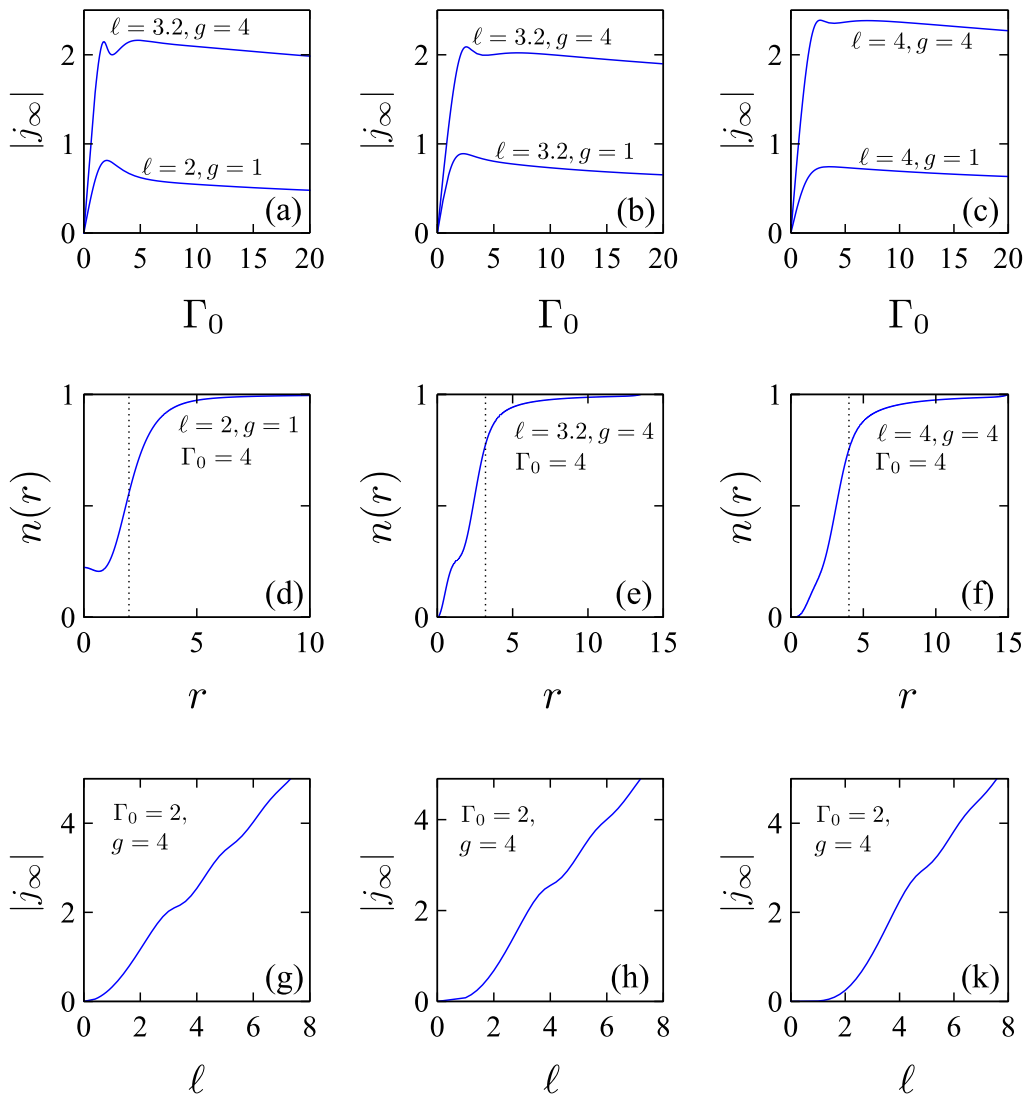}%
  \caption{ (Color online)  Dependencies $|j_\infty|$  {\it vs.}    $\Gamma_0$ for $q=0$ (a),   $q=1$ (b), $q=2$ (c) obtained for fixed $\ell$ and $g$.
     Densities  $n(r)$ of   particular stationary flows  from panels (a)--(c) are shown in (d)--(f). Vertical dotted lines correspond to the width of the dissipation, $r=\ell$.
   (g)--(k)  Dependencies $|j_\infty|$  {\it vs.}    $\ell$ for $q=0$ (g),   $q=1$ (h), $q=2$ (k) obtained for   fixed $\Gamma_0$ and $g$. In all the panels, the specific values of parameters $\ell$, $\Gamma_0$ and $g$ are labelled in the plots (where applicable).}
  \label{fig-j(g)m=0}
\end{figure}

Passing now to numerical results, we  considered stationary flows with zero vorticity ($q=0$), as well as singly-  ($q=1$) and doubly-quantized ($q=2$) vortex flows.  We numerically identified stationary flows for different representative sets of the model parameters. The outcomes of our study are summarized in Fig.~\ref{fig-j(g)m=0}.
  In the upper row [Fig.~\ref{fig-j(g)m=0}(a)--(c)],  we present dependencies of the absolute value of the radial current  density $|j_\infty|$ on $\Gamma_0$  found for a fixed width of the dissipation $\ell$ and the strength of nonlinearity $g$ (notice that $j_\infty$ is always negative which follows from (\ref{eq:ODEsysb})). We observe that the  dependencies $ j_\infty(\Gamma_0) $ are  non-monotonous, i.e. they feature one (or several) local   maxima, and   the current density  $|j_\infty|$ decreases for sufficiently large $\Gamma_0$, which means that lower current density is necessary to compensate the effect of stronger dissipation. The observed nonmonotonicity is   the evidence of the macroscopic Zeno effect. The effect is pronounced  for the flows with either zero or nonzero  vorticity $q$. Bearing in mind that the atomic losses in the dissipative domain are proportional to the obtained current densities, we observe remarkable qualitative similarity of the obtained curves with the experimentally observed ionization rate of condensed atoms subjected to the electronic beam (see Fig.~2 in \cite{Ott-experiment}).

\begin{figure}
\includegraphics[width=0.75\textwidth]{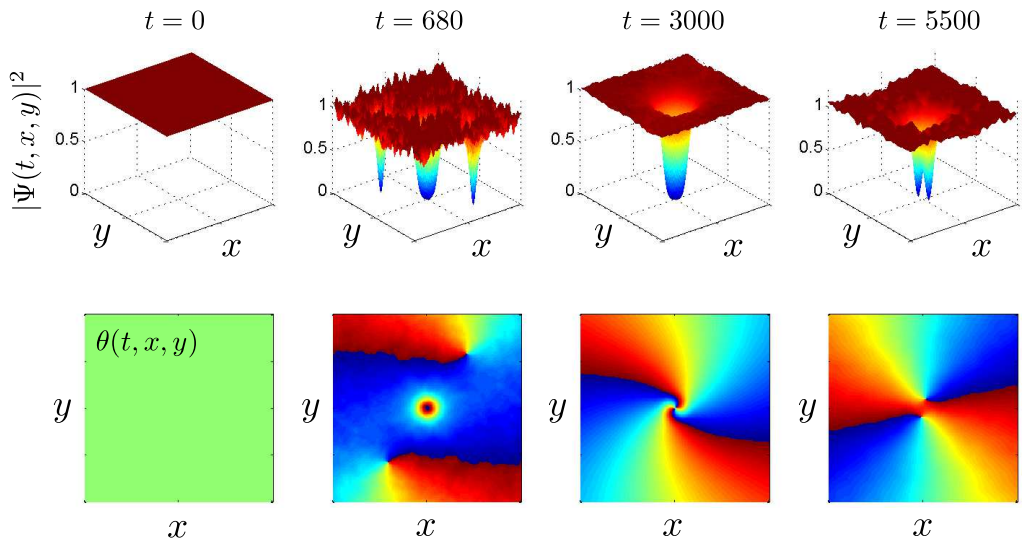}%
  \caption{(Color online) Formation (at $t\leq 3000$) and destruction (at $t>3000$) of a doubly-quantized stationary flow. Parameters of the model:  $\Gamma_0=5$ for  $t\leq 3000$ ( $\Gamma_0=0$ for   $t > 3000$), $\ell=4$,   $g=1$. Upper row shows density of the BEC $|\Psi(t, x, y)|^2$, while  the lower row shows the phase $\theta=\arg\Psi$. All the panels are presented in the spatial window $(x,y)\in [-20,20]\times [-20,20]$.}
  \label{fig-dyn}
\end{figure}

Typical spatial profiles of stationary density distribution $n(r)$   are shown in   panels  Fig.~\ref{fig-j(g)m=0}(d)--(f). It is interesting to notice that   the density of a stationary flow with zero vorticity $q=0$  [panel (d)] features   a  local maximum  exactly at the maximum of  the  dissipation, i.e. at $r=0$ (a similar effect was   also   observed in 1D case~\cite{ZKBO}).

We have also considered dependence of the current density $j_\infty$ on the width of the defect $\ell$ for a given $\Gamma_0$ and $g$, Fig.~\ref{fig-j(g)m=0}(g)--(k). For all the considered  cases,  these dependencies were    found to be monotonous, which contrasts   the situation   reported in~\cite{ZKBO} for the 1D case.

Using Eqs. (\ref{eq:LinStabMatrix}),  we examined linear stability of the flows  shown on Fig.~\ref{fig-j(g)m=0} by computing the eigenvalues $\omega$ for
$\kappa=0,1,\ldots 5$. Our  analysis  indicated that the stationary flows shown on Fig.~\ref{fig-j(g)m=0} are \textit{stable}.  We   also confirmed stability of the presented   stationary flows by means of computing their temporal evolution, which pointed out  that the solutions  are robust against   relatively small perturbations introduced  into the initial conditions. Moreover, we observed that the stationary flows feature attractor-like behavior, that is they can be obtained starting from a fairly general class of   initial conditions. This fact is illustrated in Fig.~\ref{fig-dyn} where the  initial conditions with   uniform density ($|\Psi(0, x,y)|^2=1$) undergo a sufficiently long transient process (see e.g. panels with $t=680$) and eventually form a doubly-charged stationary flow ($t=3000$). The resulting vorticity  ($q=2$ in Fig.~\ref{fig-dyn}) is determined by the boundary conditions which  are chosen to ``swirl'' the condensate. The boundary conditions also fix the density of the condensate at the infinity ($|\Psi|^2=1$), but do not prescribe any specific value of the   flux $j_\infty$ incoming from the infinity. However, after the transient process the persistent incoming flux is established automatically, and the structure shown in  Fig.~\ref{fig-dyn} at $t=3000$  is stationary and stable, i.e., it can  persist  for indefinitely long time, provided that the parameters of the model do not change. However, in the numerical experiment illustrated in Fig.~\ref{fig-dyn}, we present yet another dynamical scenario when,   after an abrupt switching off the dissipation  ($\Gamma_0=0$ for  $t>3000$), the obtained structure breaks into a pair of singly-charged vortices.  In general,   the GPE without   dissipation [i.e., Eq.~(\ref{GPE}) with $\Gamma(\br)=0$ which is tantamount to the defocusing nonlinear Schr\"odinger equation] is known to possess long-living doubly-charged vortex solutions \cite{Aranson}. In our case, however,  the stationary density distribution at $t=3000$ and $\Gamma_0=5$ is sufficiently different from that of the double vortex  supported by the GPE without the dissipation so that no formation of the conservative  doubly-charged occurs, but the density splits into two single-charged vortices.


\section{Conclusion}

To conclude, we have reported the existence of the macroscopic Zeno effect in a two-dimensional Bose-Einstein condensate subjected to a radially symmetric localized dissipation. The effect manifests itself in a non-monotonic dependence  of the superfluid current density, necessary for supporting stable stationary flows, on the strength of the dissipation. When the latter is relatively small,  its increase results in increase of the  current density. However, after a certain threshold value   further increase of the strength of the dissipation requires lower currents necessary to compensate the losses. The phenomenon is originated by a dissipation, which must have sufficiently strong localization (or even finite support, as in our case) but nonzero width (no Zeno effect can be observed for a point-like dissipation).

We have demonstrated that the effect can be observed for solutions that bear either zero or nonzero topological charge. In the latter case, the solutions represent stationary vortex flows. We found that such solutions are dynamically stable, either for single or multiple topological charge.

Our study was focused on an application to an atomic  Bose-Einstein condensate. In the meantime, mathematical analogy allows one to predict a variety of other physical systems, such as  optical waveguides, exciton-polariton  and magnon condensates (see the discussion in~\cite{ZKBO}), where observation of macroscopic Zeno effect is also possible.

\acknowledgments
 
The authors greatly acknowledge stimulating discussions with H. Ott and G. Barontini. The work was supported by the FCT  (Portugal) through the grants PEst-OE/FIS/UI0618/2014 and PTDC/FIS-OPT/1918/2012.

\end{document}